\documentclass{article}
\usepackage{authblk}
\usepackage[utf8]{inputenc}
\usepackage{todonotes}
\usepackage{booktabs}
\usepackage[hyphens]{url}
\usepackage{subcaption}
\usepackage{graphicx}

\begin{document}
\title{Analysis and Correlation of Visual Evidence in Campaigns of Malicious Office Documents}
\author[1,2]{Fran Casino}
\author[3]{Nikolaos Totosis}
\author[1]{Theodoros Apostolopoulos}
\affil[1]{Department of Informatics, University Piraeus, 80 Karaoli \& Dimitriou str, 18534 Piraeus, Greece}
\author[1]{Nikolaos Lykousas}
\author[1,2]{Constantinos Patsakis}
\affil[2]{Information Management Systems Institute of Athena Research Center, Greece}
\affil[3]{Hatching, Netherlands}
\date{}
\maketitle
\begin{abstract}
    Many malware campaigns use Microsoft (MS) Office documents as droppers to download and execute their malicious payload. Such campaigns often use these documents because MS Office is installed in billions of devices and that these files allow the execution of arbitrary VBA code. Recent versions of MS Office prevent the automatic execution of VBA macros, so malware authors try to convince users into enabling the content via images that, e.g. forge system or technical errors.

    In this work, we leverage these visual elements to construct lightweight malware signatures that can be applied with minimal effort. We test and validate our approach using an extensive database of malware samples and identify correlations between different campaigns that illustrate that some campaigns are either using the same tools or that there is some collaboration between them.
\end{abstract}

\section{Introduction}
While cybercrime has always been a threat, it has evolved into a multi-billion underground economy over the past few years. The economic impact of cybercrime \cite{crime_report,thomas2020cybercrime} is so is devastating that according to the World Economic Forum considers it the second most-concerning risk for global commerce over the next decade \cite{wef}. Moreover, the recent COVID-19 pandemic and the spike in usage of digital services has also resulted in an analogous increase of cybercrime activities as reported by multiple sources\footnote{\url{https://www.europol.europa.eu/newsroom/news/covid-19-sparks-upward-trend-in-cybercrime} \url{https://www.interpol.int/en/News-and-Events/News/2020/INTERPOL-report-shows-alarming-rate-of-cyberattacks-during-COVID-19}}.

As cybercrime evolves in terms of scale and sophistication, artificial intelligence (AI) helps resource-intensive security operations by using technologies such as machine learning, pattern recognition, and natural language processing, which are capable of ingesting terabytes of unstructured data to enhance response times and expand the capacities of security operations. Nevertheless, attackers tend to be a step ahead due to the continuous appearance of novel technologies, industrialisation processes, the difficulty to collect data from different sources in orchestrated campaigns and their timely detection, and the lack of proactive security mechanisms. Undoubtedly, beyond the underground economy exchange (drugs, trafficking etc.), a significant share of this impact stems from the exploitation of security issues that allow an adversary to monetise vulnerabilities, e.g. by injecting commands and manipulating a compromised system or network traffic, by performing extortions, etc. Pervasive and sustained cyber-attacks could have a potentially devastating impact on national and international organisations, disrupting the operations of governments and businesses and the lives of private individuals.

One of the most used cybercrime-related activities is phishing as it can be used to launch a series of attacks to the victims. The adversary tries to exploit the human factor by presenting an email that looks benign, e.g. appears to originate from a trusted source or having a harmless attachment; however, the attachment has a malicious payload. While attaching an executable may provide the adversary with immediate access to the victim's machine, this method is not used a lot. The reason is that executables are most often blocked from mail servers and that users do not usually receive such content via email. Therefore, the victim is less likely to receive it and open it. On the contrary, an email containing a Microsoft (MS) Office Document, e.g. a Word document or an Excel spreadsheet is more likely to be opened.

\noindent \textbf{Motivation and Contribution:}
Malware packed in MS Office documents was quite common in the past as the embedded macros were automatically executed when the corresponding trigger is launched, e.g. document is opened/closed etc. To address this issue, recent versions of MS Office have macros disabled by default reducing the success rate of such attacks. Nevertheless, the problem is not by any chance solved as repeatedly shown by the impact of the associated malicious campaigns.

\begin{figure}[th]
    \centering
    \includegraphics[trim={0 1cm 0 1cm},clip,width=\textwidth]{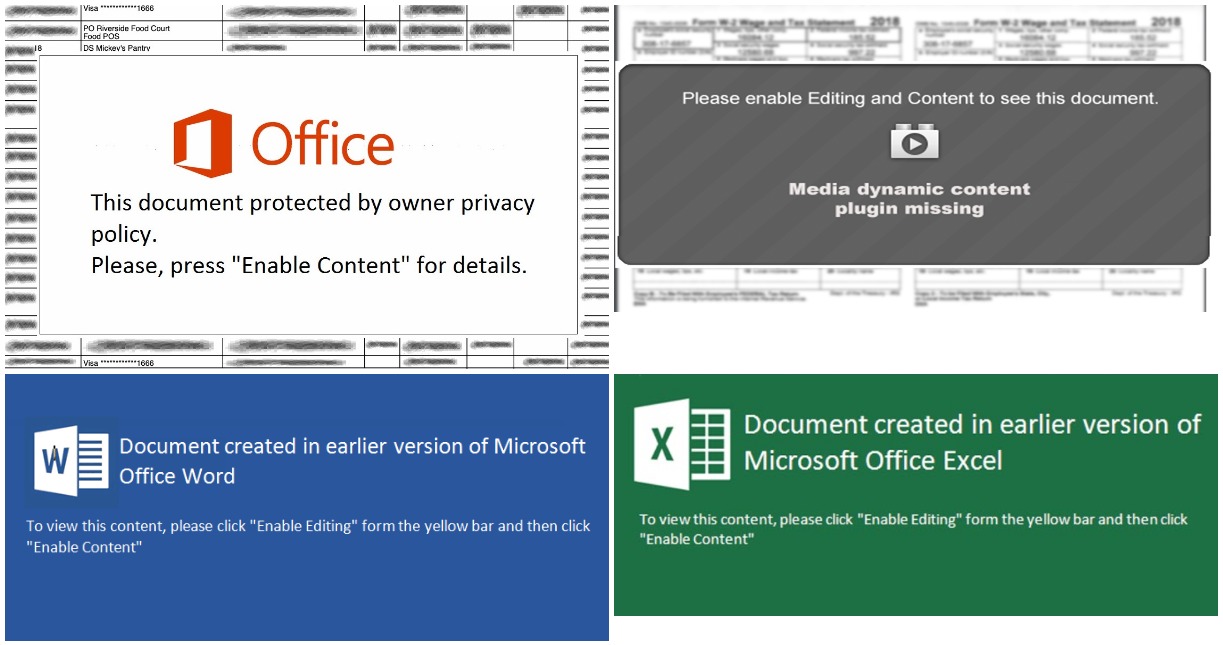}
    \caption{Deceptive images in weaponized MS Office documents to convince the user to enable content.}
    \label{fig:enable_content}
\end{figure}

\begin{figure}
    \centering
    \includegraphics[width=\textwidth]{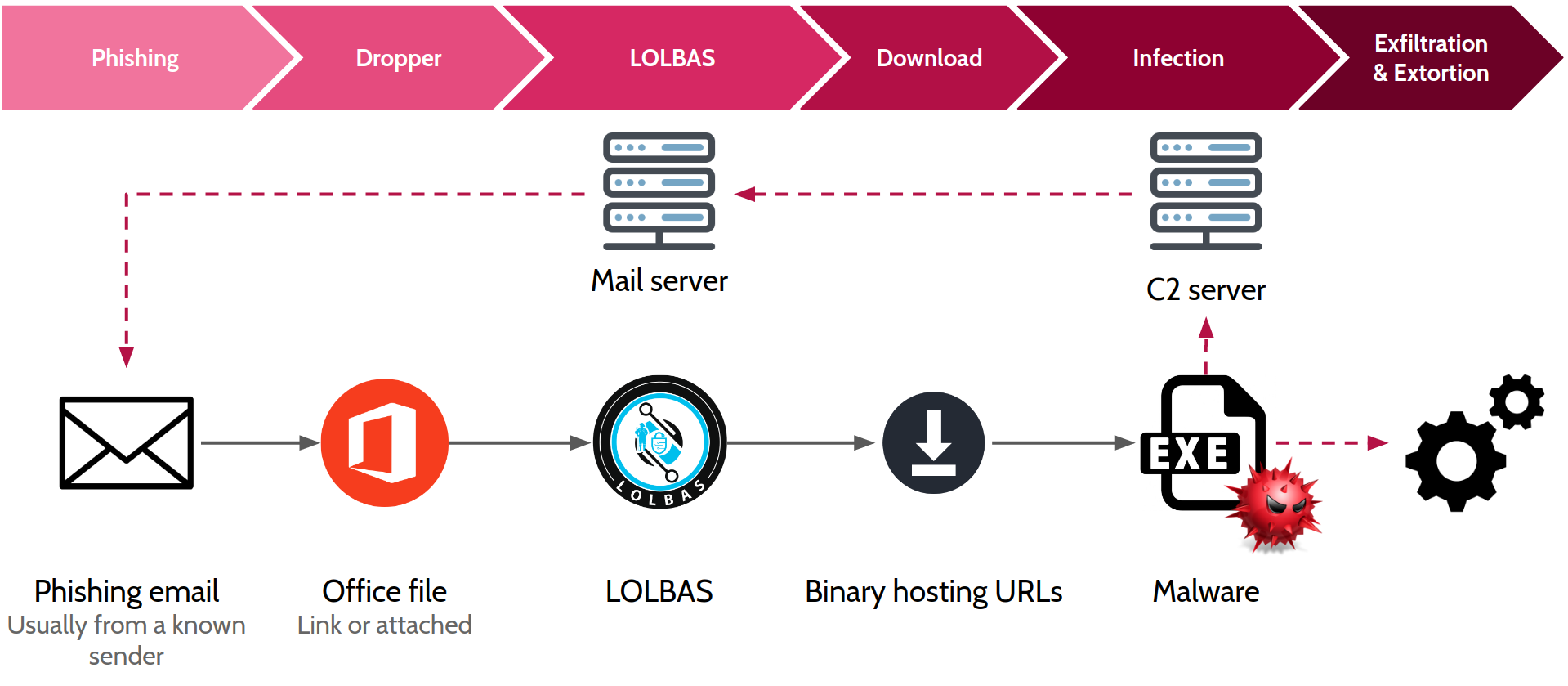}
    \caption{Modus operandi of malware campaigns based on Microsoft Office documents.}
    \label{fig:modus_operandi}
\end{figure}

In most malware campaigns that are based on malicious MS Office documents, the \textit{mondus operandi} is quite typical. The adversary tries to trick the user into opening an MS Office Document that comes in an attachment or a link. While the exploitation of Dynamic Data Exchange (DDE) may offer automatic code execution, in most campaigns, the malware authors opt for macros as several patches prevent DDE execution. Nonetheless, this choice requires the victim to accept the macro's execution in her device, as in most cases, this is disabled by default. Therefore, the adversary has embedded an image in the document, which is the only thing that the user sees and tries to mislead the victim and convince it to enable the content. The image in the bulk of the cases falsely states that either some technical error has occurred or the document's data is not accessible, and only by enabling the content it can be resolved, see Figure \ref{fig:enable_content}. Should the victim be tricked into enabling the content, a macro is activated which executes a malicious payload (either contained in the file or more often downloaded) using some Living Off The Land Binaries and Scripts (LOLBAS)\footnote{\url{https://lolbas-project.github.io/}}. Upon infection, the executed malware may proceed to its core operation. The steps described before are depicted in Figure \ref{fig:modus_operandi}.

Based on the above, it is clear that one of the key components in the attack is the image that is displayed to the victim. This work aims to investigate the possible correlations between malware campaigns based on this visual piece of evidence. To this end, we have collected a big and broad dataset which consists of more than 11 thousands of malicious documents from 16 malicious campaigns. We extracted the embedded images, and we leverage them to cluster the documents per malware and campaign. We argue that this approach may act as an indicator of compromise (IOC) and prevent many attacks from the mail server or the end-point, depending on where the mechanism is deployed. Thus, contrary to the state of the art methods that are based on natural language processing (see Section \ref{sec:maldocs}), we perform more lightweight calculations, e.g. compute the perceptual hash of an image and only detect the presence of VBA, XLM macros\footnote{\url{https://support.microsoft.com/en-us/office/working-with-excel-4-0-macros-ba8924d4-e157-4bb2-8d76-2c07ff02e0b8}}, p-code, or DDE to determine whether a file belongs to known malware campaign. If the image is not known, we analyse the image to extract relevant text which is known to be associated with malicious campaigns and flag it appropriately. As a result, we create a two-layer lightweight filter which identifies malicious MS Office documents efficiently and with high accuracy.

The rest of the manuscript is structured as follows. In the next section, we present the related work, giving an overview of the structure of MS Office document files, methods used to weaponise them and to detect them. Moreover, we present the core idea of perceptual hashing, which will be used in our work. In Section \ref{sec:methodology}, we discuss the concept behind our approach and detail our methodology. Then, in Section \ref{sec:experimental}, we present our experimental results. In Section \ref{sec:discussion}, we discuss the outcomes of our experiments and our findings. Finally, the manuscript concludes summarising our contributions.

\section{Related Work}
\label{sec:related}
\subsection{MS Office document files}
\label{sec:msdocs}
\begin{figure}[th]
    \centering
    \includegraphics[width=\textwidth]{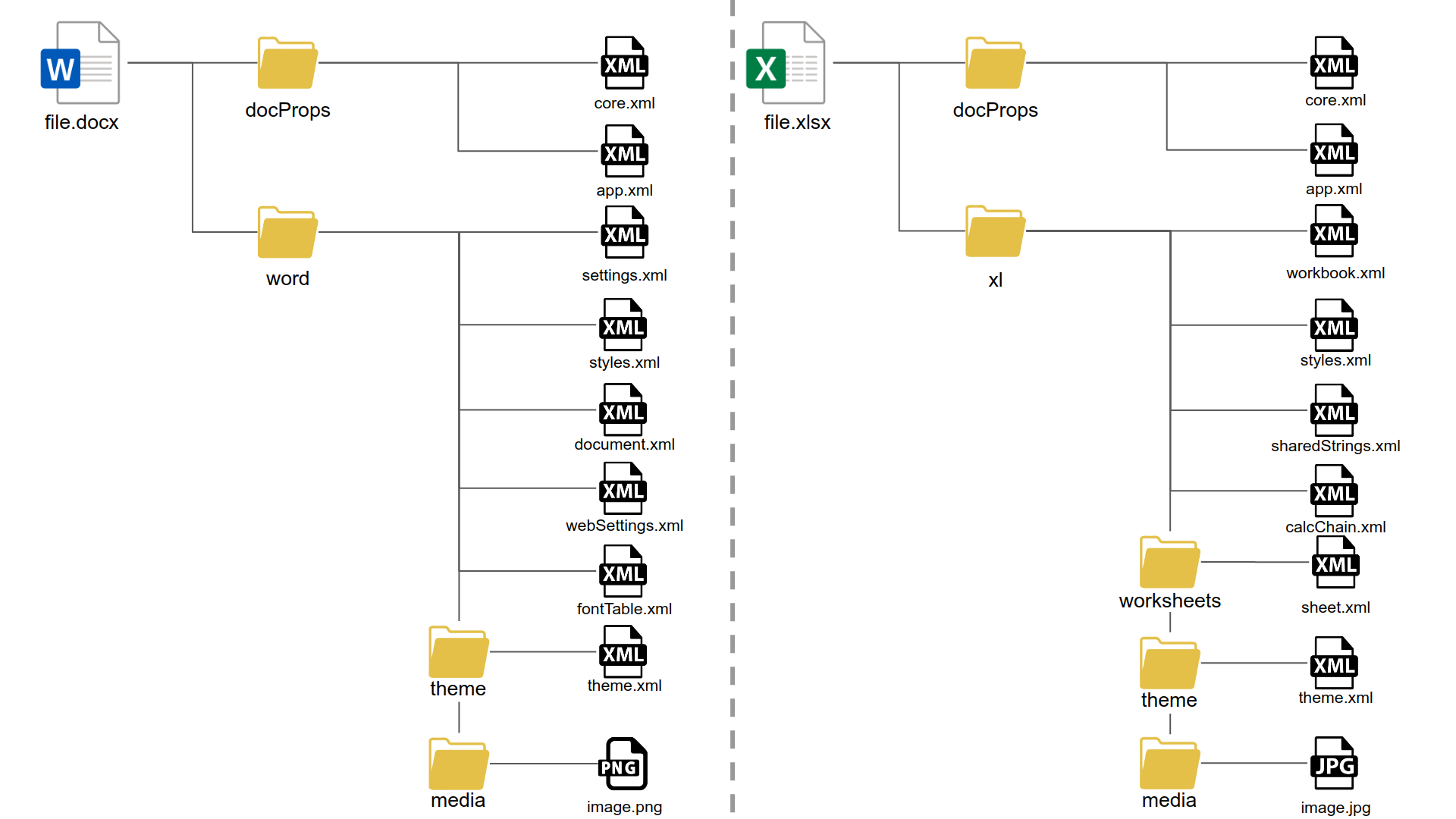}
    \caption{Typical structures of a document and a spreadsheet OOXML files.}
    \label{fig:ooxml}
\end{figure}

MS Office is considered the default suite for writing documents, working with spreadsheets, and making presentations. To allow interoperability, MS has adopted the \textit{Office Open XML}, also known as OpenXML or OOXML, format. As the name implies, it is an XML-based format for all office documents. The specification was developed by MS, has been adopted by ECMA International as (ECMA-376) \cite{ecma} and became an ISO and IEC standard (ISO/IEC 29500) \cite{iso}. In principle, all OOXML files are stored in a compressed ZIP file. Therefore, each office file contains a set of XML files and stores the necessary files along with the schema. Images, audio or other multimedia files, as well as scripts stored in the document are also stored inside the same ZIP file. The typical structure of an OOXML file is illustrated in Figure \ref{fig:ooxml}.

Notably, several other formats are supported by MS Office for output, such as the Compound File Binary Format (CFBF) which are structured storage files\footnote{\url{https://docs.microsoft.com/en-gb/windows/win32/stg/compound-files?redirectedfrom=MSDN}}. In this regard, there are XLSB files supported by MS Excel, and due to their binary format they are much faster than traditional XLS/XLSX files. The latter introduces a very interesting twist as files can be encrypted. Therefore, the content cannot be directly examined by an intermediate security mechanism. While Emotet has recently used encrypted ZIP files in its campaign \cite{emotet_article}, in this scenario, there is no need for a ZIP file as the office file is directly encrypted. In fact, this approach was recently used in several recent malware campaigns which were further facilitated by an old Excel bug \cite{sweat}. This bug allows Excel to automatically decrypt an Excel spreadsheet if the used password is \textit{VelvetSweatshop}. Thus, the file was encrypted with this password, the security mechanisms could not scan the file, yet the file opened seamlessly in the user's device.

\subsection{Malicious MS Office files and their detection}
\label{sec:maldocs}
MS Office documents are weaponised in a very straightforward way. Since most MS office support VBA code for macros and XLM macros, the most commonly used method to create a malicious document is add a macro to the document which is executed automatically, e.g. upon opening or closing a workbook or document with \texttt{Workbook\_Open()}, \texttt{AutoOpen()}, and \texttt{AutoClose()} functions. The code can then use the Shell command to execute a shell command, or even download data from the Internet. Therefore, malicious MS Office documents are often used as droppers, that is they download malicious binaries from the Internet, and they execute them, passing the control to them. Their authors obfuscate their code by adding unused code, base 64, hex and octal encoding, break strings into smaller ones or results of functions, even abuse MS Office related functions to prevent their analysis.

A method to hide VBA code's execution is to actually destroy the VBA source code in the document but leave the compiled version of the macro code known as p-code. This method, known as VBA stomping, originally by V. Bontchev\footnote{\url{https://github.com/bontchev/pcodedmp}}, bypasses static analysers which try to simply extract the VBA code from a document, however, Office will execute the payload from the p-code \cite{stomp}. Finally, one may inject a malicious payload using Dynamic Data Exchange (DDE) exploiting several MS Office vulnerabilities linked to it, such as CVE-2017-8759, CVE-2017-11292, and CVE-2017-11826, or use XLM macros.

To detect malicious office documents, many researchers are using natural language processing methods \cite{kim2018obfuscated,yamin2018detecting,mimura2019using,mimura2019towards} to detect the presence of obfuscated code in VBA macros which is linked with the document being malicious. Due to the imbalances in the available datasets with such documents to train machine learning algorithms to detect such files, Mimura \cite{mimura2020using} recently proposed a method using Generative Adversarial Networks to generates fake samples with similar properties.

In another research line, researchers try to exploit n-grams \cite{bearden2017automated} of the documents or other similar features such as entropy and other byte-level statistics over fragments of the data stream \cite{rudd2018meade}. For more details on the threats from malicious documents \cite{257190} the detection of malicious documents, the interested reader may refer to \cite{maldoc_survey}.

\subsection{Perceptual hashing}
\label{sec:perceptual}
Traditionally, hash functions are used to create an easy way to deterministically collect a small ``sample'' from a data stream that can be used to identify it and differentiates it from others with overwhelming probability. Nevertheless, we may tolerate small variations for images as long as the actual content remains the same. In this regard, we want to create a hash that remains the same after small image manipulations, e.g. rotation, cropping, or some light colour distortion. This type of hashing is called perceptual hashing, and it is widely used to facilitate tasks such as image search, retrieval, and authentication. Therefore, perceptual hashing is ideal for finding similar images.

\begin{figure}
    \centering
    \includegraphics[width=\textwidth]{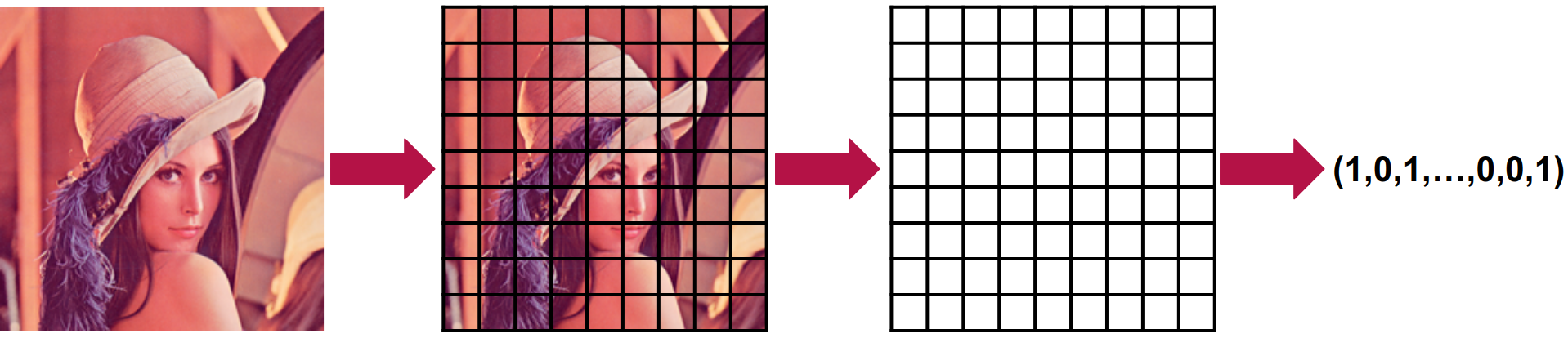}
    \caption{Overview of how perceptual hashing works.}
    \label{fig:perceptual}
\end{figure}

Generally, we split the image on a grid and extract features from each segment. These are then transformed into a vector which is translated into the perceptual hash of the image. There are various methods for perceptual hashing, which in general fall into five main categories. According to Du et al. \cite{du2020perceptual} these are Invariant feature transform-based methods, Local feature points based methods, dimension reduction based methods, Statistics features based methods, and Learning based methods.

\section{Methodology}
\label{sec:methodology}
In what follows, we present the basic concept of our methodology and then our approach, consisting of our methodology and the collected dataset.
\subsection{Concept}
\label{sec:concept}
In many attacks, the adversary tries to create an asymmetry in the balance between his needed effort to launch an attack and the victim's effort to detect and mitigate it. For instance, in the scenarios that we are investigating, the attacker tries to create many documents with small variations on the VBA code, some metadata, as well as text, which will lead to different hashes of the attachment. This way, one cannot simply blacklist a specific hash of a file or some code fragment.

While we understand that the latter task can be easily achieved by an adversary which already has many tools to achieve this, we argue that the same does not apply for the visual part of the document. For the execution of the VBA, the attacker has to convince the victim to enable the content. Therefore, random images, not clear, and without visual guidance to enable content would significantly decrease the attack's success. Based on the above, we argue that by detecting the visual part of the attack, we increase the adversary's effort, and we may create much more efficient filters. As a result, we aim to extract the images that are contained in a document or a spreadsheet and correlate the information to i) create more efficient IoCs, ii) identify the use of common tools as well as cooperation between different malicious actors and campaigns. Therefore, our hypotheses are the following:
\begin{enumerate}
    \item The images contained in an MS Office document can be a good indicator that a file is malicious.
    \item Malicious campaigns and threat actors reuse images in campaigns due to the use of the same tools as well as cooperation.
\end{enumerate}

\subsection{Our Approach}
\label{sec:approach}
To assess the applicability and efficacy of our hypotheses we collected a large dataset of malicious files from Triage and Malware Bazaar. The office documents belong to 16 different campaigns, as illustrated in Table \ref{tbl:families}.
\begin{table}[ht]
    \centering
    \begin{tabular}{lr}
    \toprule
         \textbf{Malware Family}&  \textbf{Samples}\\
         \midrule
            AgentTesla &16\\
            Dridex & 312 \\
            Emotet &4,268\\
            formbook &124\\
            Hancitor & 35 \\
            IcedID & 505 \\
            Loki & 63 \\
            masslogger & 26\\
            Netwire & 45\\
            Qbot & 3,699\\
            Remcos & 14\\
            Smokeloader & 998\\
            TaurusStealer & 30\\
            TrickBot & 55 \\
            Gozi\_Isfb & 212\\
            ZLoader & 729\\
         \midrule
         \textbf{Total} & \textbf{11,131}\\
         \bottomrule
    \end{tabular}
    \caption{Composition of our dataset as of samples per family.}
    \label{tbl:families}
\end{table}

After collecting these samples, we exploited the fact that most office files are actually ZIP files so the stored multimedia can be easily exported without tampering the files and without opening the files in any virtual machine. Since several of the samples were XLSB files encrypted with the \textit{VelvetSweatshop} password (see Section \ref{sec:msdocs}), we utilised msoffcrypto-tool\footnote{\url{https://github.com/nolze/msoffcrypto-tool}} to decrypt the content and create a typical Excel file, without distorting its contents. This method is very efficient and lightweight; it guarantees that we are extracting the artefacts in a forensically secure manner and that no malicious code can be executed from the sample. Once the images have been collected from each sample, we compute its hash (using SHA-256), the perceptual hash of the image, and then proceed with the text of the image. To this end, we detect the text's language, extract the text of the image, and translate it, where applicable. The extracted information is stored in a database and used for correlating, assessing, and clustering, as discussed in the following paragraphs.

\begin{figure}[th]
    \centering
    \includegraphics[width=\textwidth]{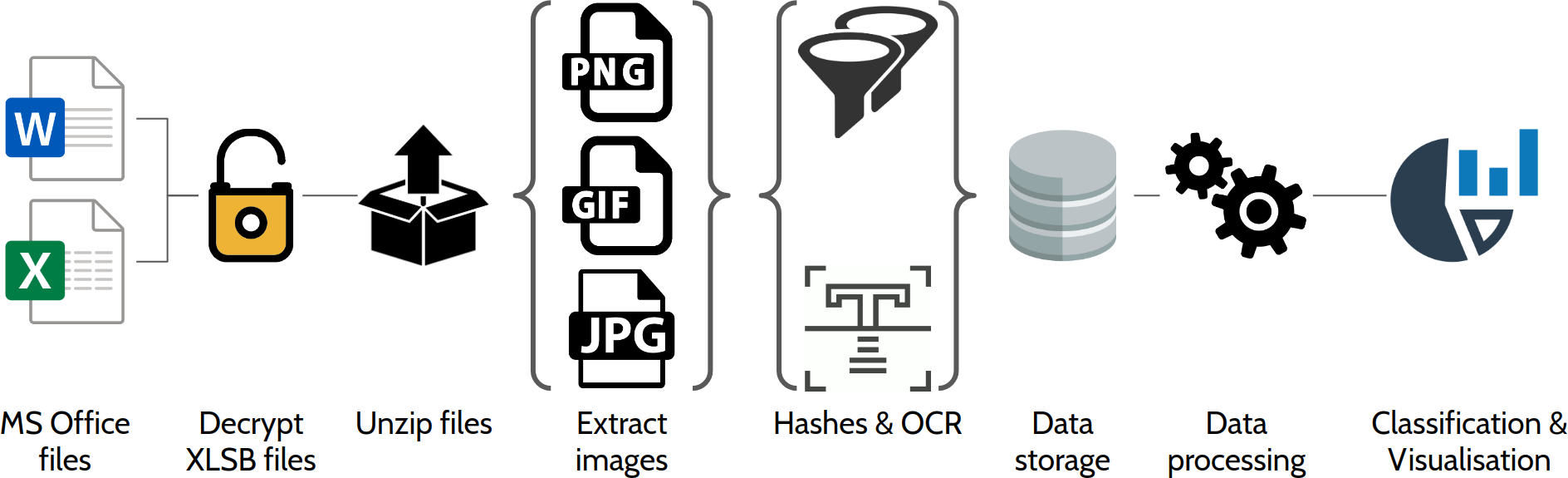}
    \caption{Overview of our methodology.}
    \label{fig:analysis}
\end{figure}


\section{Experimental results}
\label{sec:experimental}

In the following sections, we provide an exploratory analysis of our dataset as well as a description of the methods used to process the data and the corresponding outcomes.

\subsection{Exploratory Analysis}

\begin{figure}[th]
    \centering
    \includegraphics[width=\textwidth]{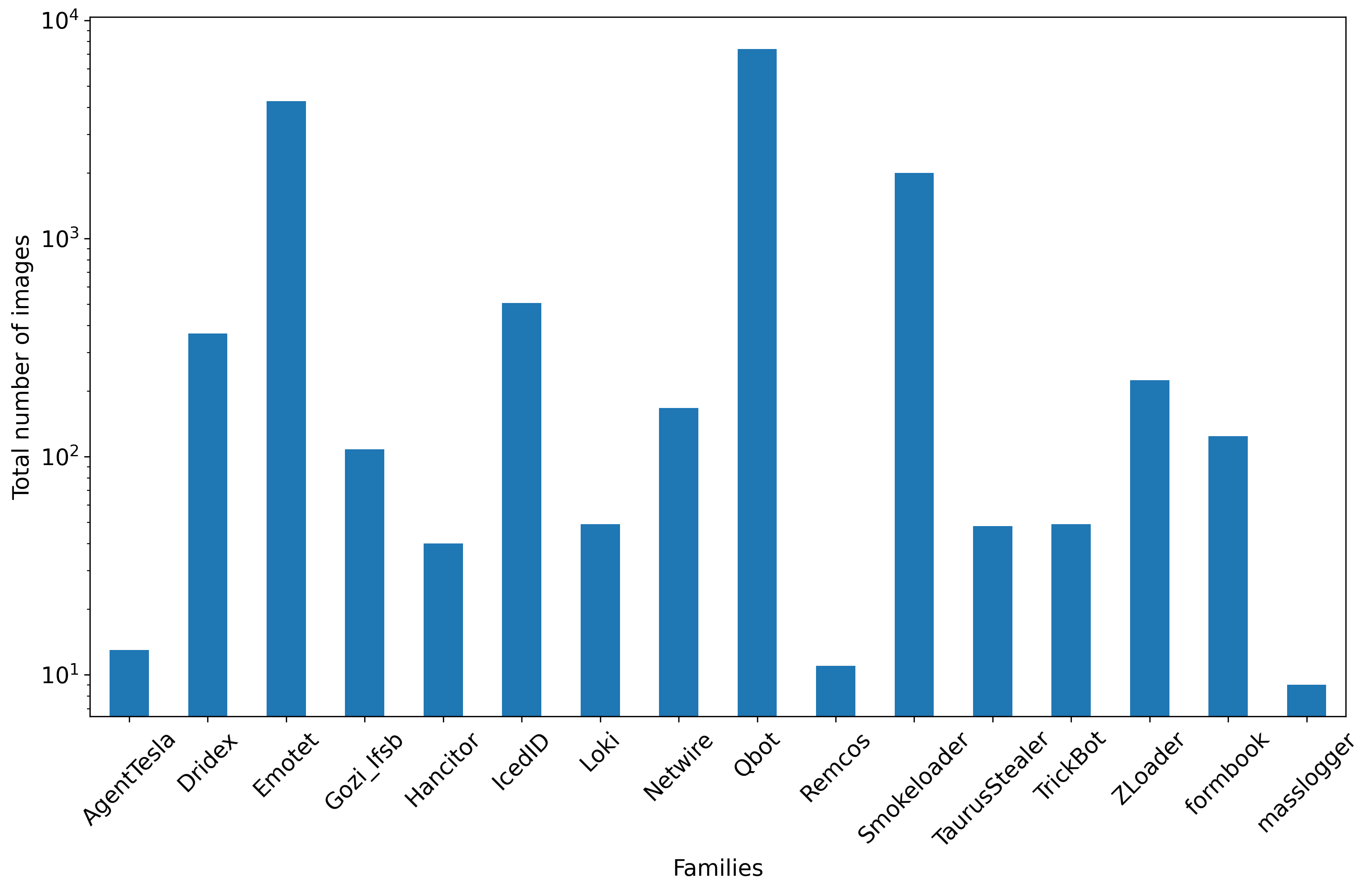}
    \caption{Number of images retrieved from each corresponding family, according to unique SHA-256 values. Values are represented in logarithmic scale. }
    \label{fig:images_freq}
\end{figure}

As described in Table \ref{tbl:families}, we collected samples from different malware families. Moreover, each of such samples contained a set of images. In this regard, Figure \ref{fig:images_freq} depicts the amount of images collected from each family. As it can be observed, Qbot, Emotet, and Smokeloader were the most populated families. It is relevant to note though that some families used more images per sample on average. For instance, Emotet has more samples than Qbot in our dataset, yet the latter exhibited a higher number of images.

\begin{figure}[th]
    \centering
    \includegraphics[width=\textwidth]{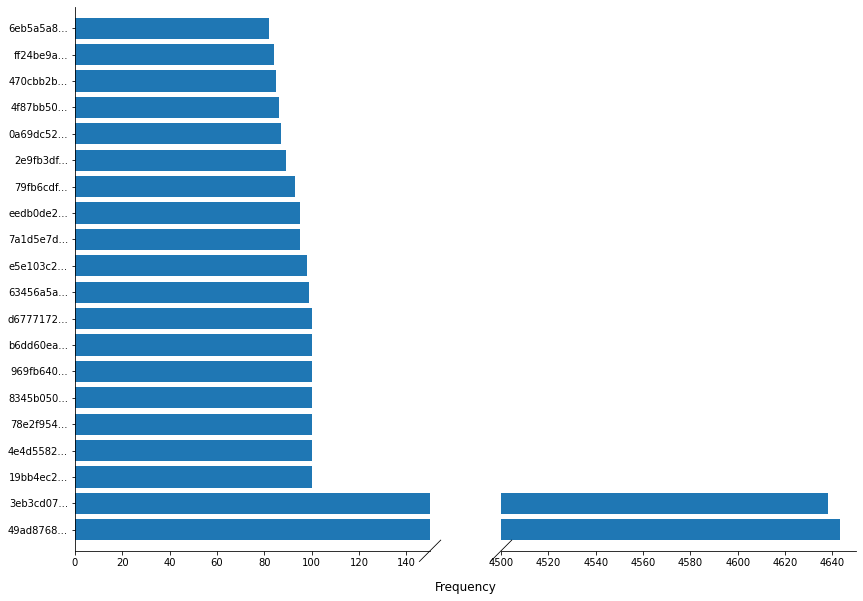}
    \caption{Top Number of images (according to their SHA-256) retrieved from each corresponding family.}
    \label{fig:sha_dist}
\end{figure}

The next experiment focused on observing how many times the same image was used in the collected samples. In this regard, Figure \ref{fig:sha_dist} shows the amount of times that an image was detected in our dataset, according to its SHA-256 hash. Clearly, there are two hashes, namely \texttt{3eb3cd078172...} and \texttt{49ad87680a...} which appeared approximately 4650 times. Such hashes belong mainly to Qbot and smoketbot, which used them exactly 3638 and 998 times, respectively. Moreover, they were used by ZLoader and Hancitor, yet only in very few samples. Therefore, these families used the same image multiple times in their campaigns. The next most used image appeared 100 times, and the pace decreases smoothly after that value. In total, we found 623 images appearing only once from the 1090 unique images collected from our samples.


\begin{figure}[th]
    \centering
    \includegraphics[width=\textwidth]{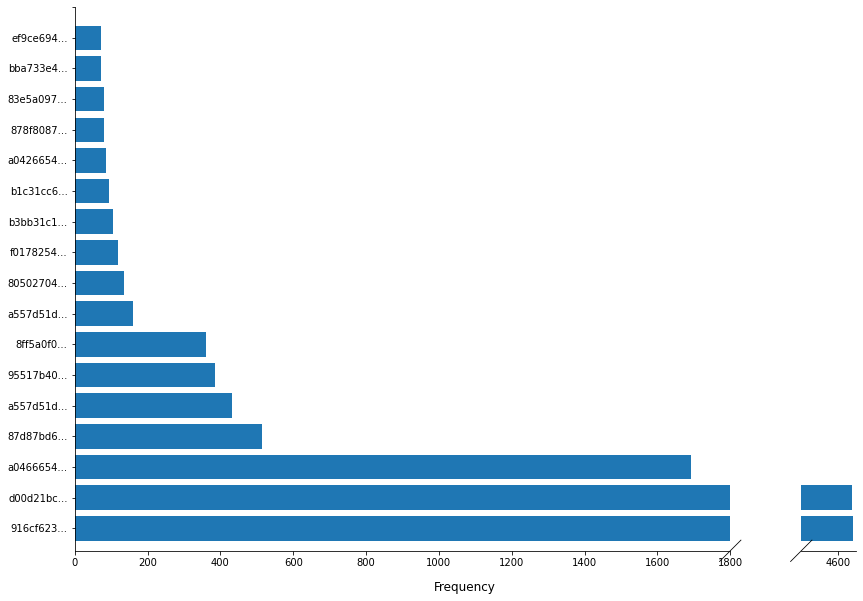}
    \caption{Top Number of images (according to their perceptual hash) retrieved from each corresponding family.}
    \label{fig:phash_dist}
\end{figure}

\begin{figure}[th]
    \centering
    \includegraphics[width=\textwidth]{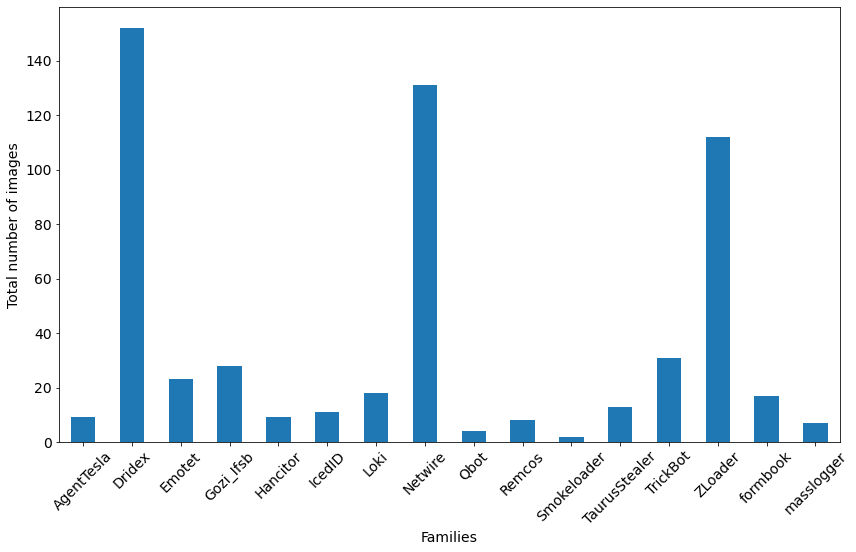}
    \caption{Number of images retrieved from each corresponding family according to unique perceptual hash values.}
    \label{fig:phash_uniq}
\end{figure}

However, the latter does not depict the actual truth as malicious actors manipulate the images to have minor distortions, most of which do not have any observable change for the human eye. This way, even if the payload is the same, the resulting file has a different hash. Therefore, to correlate the collected images and bypass the slight modifications, we computed their perceptual hashes and leveraged the same experiments that we previously performed for the SHA-256 hashes. In this regard, Figure \ref{fig:phash_dist} shows the amount of images and their appearance in our dataset. As it can be observed, many of them appear more than 100 times, contrary to the behaviour depicted in Figure \ref{fig:images_freq}. Moreover, the amount of perceptual hashes appearing only once is 304 from a total of 526 unique perceptual hashes, which also showcases the strong similarity of some images and supports the use of perceptual hash in this context. In addition, we also depicted the number of images per family, according to their perceptual hash in Figure \ref{fig:phash_uniq}.

From the comparison between the statistics shown in Figure \ref{fig:images_freq} and the ones represented in Figure \ref{fig:phash_uniq}, it is evident that several families used the same images across different campaigns due to the dramatic reduction of the number of images when removing duplicates. Yet, the set of unique images differs substantially between families. For instance, a considerable amount of Qbot, Smokeloader and Emotet samples were processed, yet such families tend to use the same subset of images in their corresponding samples, a fact which is evinced in Figures \ref{fig:phash_uniq} and \ref{fig:phash_dist}, and also by the amount of images appearing only once (i.e. from 623 to 304).

In the next paragraphs, we will explore the connections between the different families and the subsets of images used by more than one family.

\subsection{Correlations between different campaigns}
\label{sec:correlations}
\begin{figure}[th]
  \centering
  \begin{subfigure}[t]{0.5\linewidth}
      \centering
      \includegraphics[width=\linewidth]{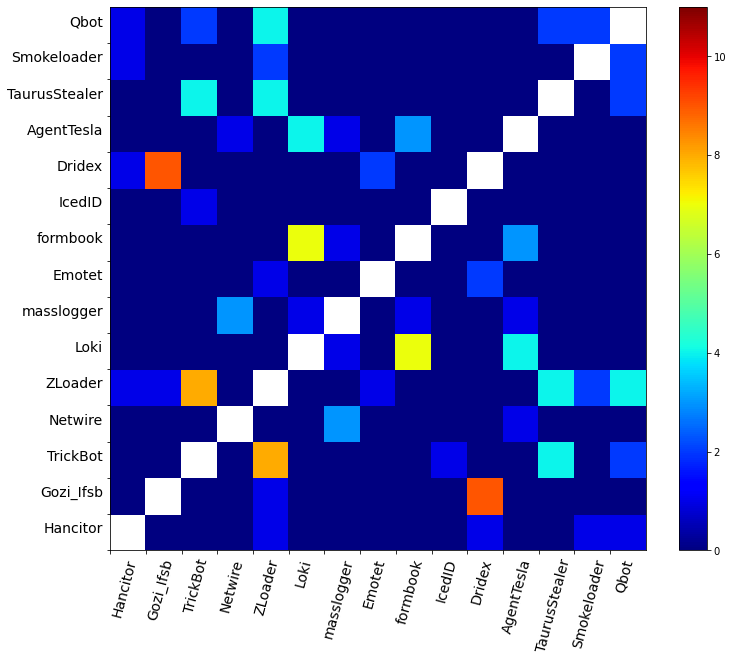}
      \caption{Heat map of the amount of unique images used by different families, according to the SHA-256 of the images.}
      \label{fig:heat_sha}
  \end{subfigure}%
  ~
  \begin{subfigure}[t]{0.5\linewidth}
      \centering
      \includegraphics[width=\linewidth]{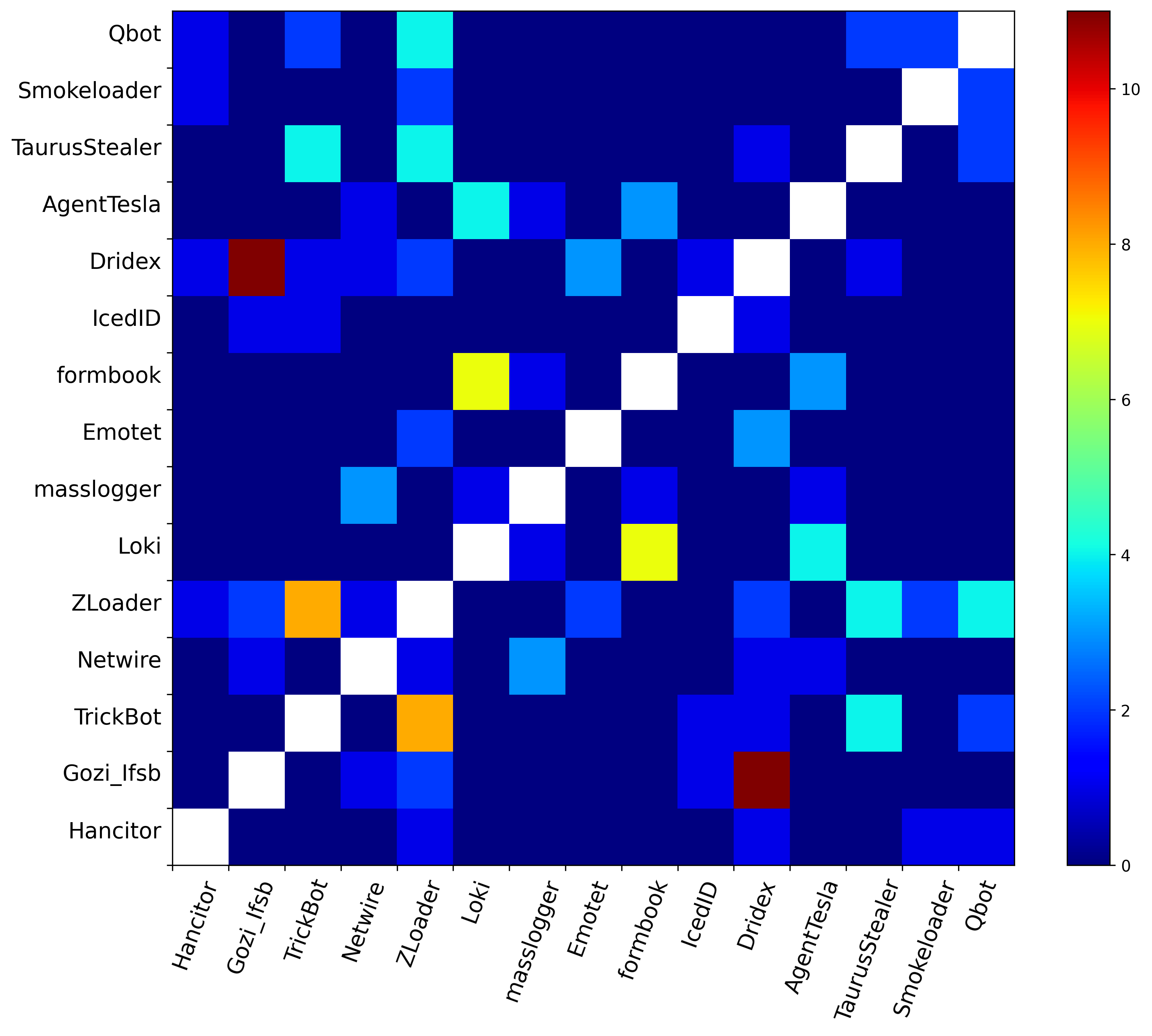}
      \caption{Heat map of the amount of unique images used by different families, according to the perceptual hash of the images}
      \label{fig:heat_phash}
  \end{subfigure}

  \caption{Cross-family interactions according to the SHA-256 and the perceptual hash of the images collected in our dataset}
  \label{fig:heatmap}
\end{figure}

Figure \ref{fig:heatmap} shows the amount of times that a unique image was used by different families. In the case of images identified by their SHA-256, we found 36 unique hashes used by more than one family, and 41 in the case of using their perceptual hash. Moreover, as it can be seen in Figure \ref{fig:heatmap}, the heat map is denser in the case of perceptual hashes (cf. Figure \ref{fig:heat_phash}), denoting that different families are using almost identical images with slight modifications, which are obviously enough to modify the SHA-256 representation of the image but not their perceptual hash. Note that some images differ only in some bits due to, e.g. a minor colour change in some pixels. Moreover, note that in Figure \ref{fig:heatmap} we do not consider the repeated use of an image by different families, which would yield much higher numbers due to the use of the same, or almost identical images in different campaigns.

According to Figure \ref{fig:heat_sha}, the families that shared the most unique images were Gozi\_Isfb and Dridex, closely followed by ZLoader and Trickbot. Moreover, Loki and formbook are also using the same subset of images in several campaigns. These correlations are strengthened when using perceptual hashes to represent the images, as observed in Figure \ref{fig:heat_phash}. In this regard, images that were slightly different were now captured and correlated between Gozi\_Isfb and Dridex. In addition, we discovered that very similar images were used by different families, showcasing the possibility that either the same perpetrators are behind different campaigns, or that malicious actors are using previously published samples and materials to enhance their malware.

\begin{figure}[th]
    \centering
    \includegraphics[width=\textwidth]{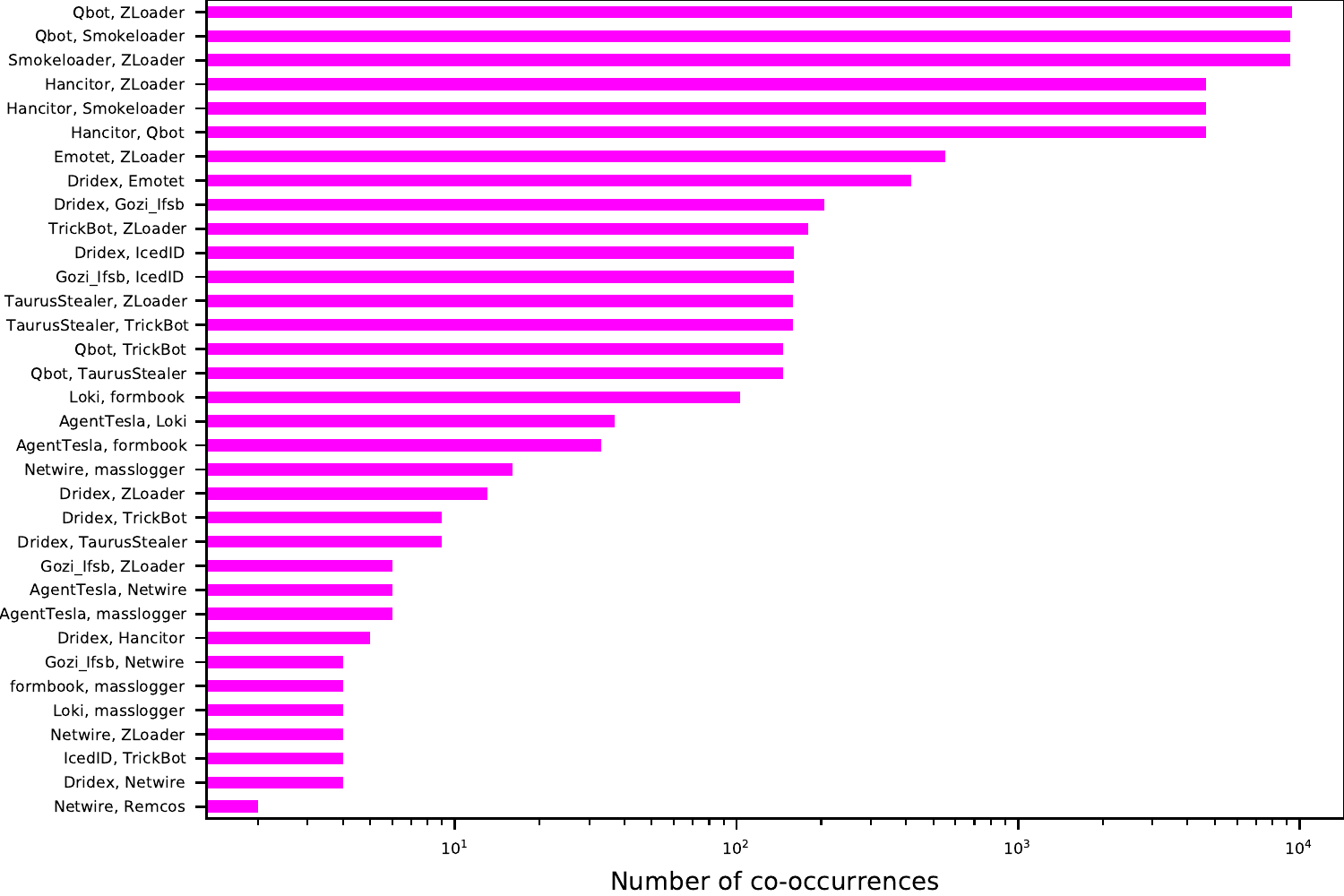}
    \caption{Co-occurrences between families considering images and samples in logarithmic scale.}
    \label{fig:collabcounts}
\end{figure}

To consider the quantity of the coincidences between different images and families, we performed another measurement. In this case, we collected the images according to their perceptual hash, searched in how many different samples they were used, and counted the co-occurrences between pairs of families. For instance, if $image_a$ appears in 10 samples of Qbot family and in 10 samples of Smokeloader family, we will count 20 co-occurrences between Qbot and Smokeloader. Note that we do not consider the direction of the co-occurrence, and thus the co-occurrences $[$Qbot, Smokeloader$]$ and $[$Smokeloader, Qbot$]$ are the same. 
The outcomes of this measurement are depicted in logarithmic scale in Figure \ref{fig:collabcounts}.

As it can be observed, there are some families atop the number of co-occurrences, namely Qbot, ZLoader, Smokeloader, and Hancitor. Therefore, we can establish a high interconnection between such families considering both the number of unique hashes (cf. Figure \ref{fig:heat_phash}), and the amount of samples in which they were used (cf. Figure \ref{fig:collabcounts}). It is worth to note that Dridex and Gozi\_Isfb did not share a high number of coincidences in this measurement despite exhibiting the highest number of unique images shared (cf. Figure \ref{fig:heat_phash}), yet the latter is directly related with the nature and amount of samples retrieved in our dataset. Therefore, to have both a global perspective of the collaborations between families, as well as the distribution of our dataset, the outcomes of both experiments need to be contrasted.

It is worth noticing that, after establishing a connection between Qbot, Zloader and Smokeloader documents and upon closer inspection, we observed that apart from the image, they shared more characteristics. More precisely, the name of their first sheet is always \texttt{DocuSign}, and they usually have two more hidden sheets. One of these sheets has the XLM macros and the other one the data used by these macros. Moreover, inspecting more samples revealed that for every different \texttt{DocuSign} image, there is a specific set of macros that accompanies it enables us to predict the behaviour of the document. The community has assigned the name \texttt{SilentBuilder} for these XLS(S|M) documents.

\subsection{Visual analysis of samples}
\label{sec:visualanalysis}
To assess the limits of our hypothesis regarding the malicious intent of images, we opted to perform a \textit{blind} test on images that are used in benign files. The blind test involves determining whether a file is benign or malicious based on the images that it contains and the fact that it has macros or DDE. While there are millions of MS documents shared online, typical users do not use macros or DDE in their documents. Therefore, public samples of such documents are very sparse. Nevertheless, we collected 890 such documents. In fact, we submitted these files to VirusTotal and Triage to validate that they are not malicious. Clearly, the aforementioned benign files introduce a bias against our methodology as typical files would not meet these requirements, nonetheless, it is the best approach to stress the efficacy of our hypothesis.

Following the methodology described in Section \ref{sec:methodology}, we extracted the images of all the collected malicious Microsoft Office samples, and then used their corresponding perceptual hash to end up with 526 unique perceptual hashes. In addition, we collected from the 890 benign samples 2497 unique images, with their corresponding perceptual hashes. Finally, we merged all these images with the malicious ones to end up with a dataset containing 3023 unique images.

To automatically detect the malicious intent of these images {\em blindly}, that is without knowing any ground truth about the file, we leverage a text detection pipeline by using Tesseract\footnote{https://github.com/tesseract-ocr/tesseract}. Therefore, we extract the text of the images and look for specific keywords.

The first step of our analysis consisted of manually annotating the images from the malicious samples that were asking the users to activate or enable content to grant access to a fully functional version of a document. Subsequently, we marked as malicious a total of 159 images. Next, we used our text pipeline extraction method, which first applies a transformation to the input image (i.e. removing the transparent layer and applying a threshold-based binarisation). Subsequently, we used Tesseract to extract the text of the image. Then we translated the text to English where applicable, to search for specific keywords such as ``enable content'' or ``enable macros''. In the event that an image contains one of the keywords tagged as malicious, we label the image as dangerous. Finally, we applied our automated method to our dataset of unique images and compared its detection rate with the manual annotations. The outcomes of such comparison are depicted in Table \ref{tbl:outcomes}. Our automated pipeline's accuracy is above 0.99, with only three false negatives, which were due to errors in the character recognition (e.g. ``enab\textbf{t}e content''). One solution for such a problem could be to, e.g. accept strings differing by one character, yet this kind of post-processing is left for future work. Note that, since the total amount of malicious files is 159, the amount of false positives has a big impact in the precision value (i.e. the amount of malicious files is around a 5.2\% of the dataset). In our experiment, we had only 11 false positives. Such images contained several of the suspicious keywords, and 10 of them were directly suggesting users enable macros, yet these images were found in benign samples.

\begin{table}[ht]
    \centering
    \begin{tabular}{cccc}
    \toprule
         \textbf{Precision}&  \textbf{Recall} &  \textbf{Accuracy} &  \textbf{F1-score}\\
         \midrule
  0.934 & 0.981 & 0.995 & 0.957 \\
         \bottomrule
    \end{tabular}
    \caption{Outcomes of our proposed method.}
    \label{tbl:outcomes}
\end{table}

One of our method's advantages is that it does not require training, so there is no need to split our dataset, and we used it as 100\% testing. Therefore, all the images were directly used in our experiment to compute the accuracy. Note that such an experiment covers the worst-case scenario, and thus, our average outcomes could reach better values if we used, e.g. subsampling or n-fold splits. Nevertheless, we wanted to state the complete numbers for the sake of clarity and to stress the efficacy of our method.

The predominant images that were found in our dataset are variants of MS Office-like text boxes asking users to enable content. Another subset of images used several combinations of colours, highlighted text and text with transparency to hinder the text detection task. A further significant subset of images is composed of images with blurred background, which creates the illusion that a real document will be unlocked if the user grants the corresponding permission. Finally, we also found other harmless images such as icons, business logos, and images belonging to step by step tutorials. It is worth noting that we found different languages corresponding to different international campaigns, including languages which used different scripts including, but not limited to, Latin, Greek, Cyrillic, Bengali, Japanese.

Based on the above, our proposed method can achieve very good results even in the case of new unknown malicious campaigns. In fact, the benign dataset that we used cannot be considered a representative of the real-world samples, as benign documents do not use macros and DDE often. Therefore, in a real-world setting, the outcomes of our \textit{blind} test would be significantly higher.



\section{Discussion and final remarks}
\label{sec:discussion}
The evolution of cybercrime into a huge underground economy has turned cybercrime into an actual industry. The above is justified by the collaboration between malware authors and the emergence of  \textit{Malware-as-a-Service (MaaS)} or \textit{Access-as-a-Service (AaaS)} models where, for instance, malware authors ``rent'' or pass the control of the compromised devices to their peers. Moreover, for many malware families, the attribution of malware to an actor is not straightforward, and due to the malware evolution and code exchange in groups, it becomes a very challenging task. For instance, Gozi has several variations with different capabilities \cite{check}, with occasional parallel campaigns of its variants. Emotet, one of the most notorious malware, is another fine example of these exchanges. It shares the same loader with Gozi\_Isfb, Dridex and BitPayme \cite{tm2018}, it has bonds with Qbot \cite{qbot}, Trickbot \cite{cybereason}, and more recently with Ryuk \cite{emo_trick,emotet_article}. The latter is supported by our experiments, since we found correlations between such families in Section \ref{sec:correlations}. More concretely, we found further correlations between families, especially in the case of, e.g. Qbot, ZLoader, Smokeloader, and Hancitor, which shared images across a high number of samples. Moreover, other families such as Dridex and Emotet also shared a relevant number of images between them and with further families such as ZLoader. Therefore, the information shared between families is higher than expected, since other families like Loki and formbook were also correlated with, e.g. Qbot and ZLoader. In summary, macro malware campaigns share more similarities than one would envision, compared with other malware campaigns.

In the aforementioned campaigns, as well as the rest included in our dataset, the first step to launch the attack is made once the user opens an MS Office document and she enables the content. If the user is not convinced to enable the content, then the attack would not start. Therefore, the malicious actors try to present a convincing message that such action is necessary as e.g. a system error occurred.

Our proposed method is very lightweight as to determine whether a file is malicious one uses a small signature which consists of the perceptual hashes of the images that it contains. If the image is in the database and the file contains VBA code, it is automatically flagged as malicious without the need to investigate the code. The latter can be easily checked with, e.g. the presence of the \texttt{vbaProject.bin} in the compressed files that comprise the document. Moreover, while the presence of obfuscated code implies that the document must be executed in a sandbox to determine in which family it belongs to, our approach can classify it far easier by the hashes. If the perceptual hash does not exist in our database, we apply the method described in Section \ref{sec:visualanalysis} to determine the threat level of the sample's images.

Perhaps the most important contribution of our work is that our approach introduces a significant effort to malware authors. The images that they use are easily flagged, and creating new and convincing ones is far from trivial. This way, even if a sample contains images which are not known, one may easily introduce them to the blacklist in the presence of specific keywords. Indeed, the latter illustrated an almost perfect efficacy. In this regard, the effort of the adversary is significantly increased. The automatic use and minor tampering of images are prevented, and the generation of convincing images cannot be automated.

Finally, our work showcases the commonalities of malicious campaigns from another perspective. The existence of so many common images among campaigns shows that either the same tools are being used or that the same people are behind them as the manipulation of the images cannot be the same for all of them. It should also be noted that some families, e.g. Emotet, are making each image unique which signifies that they are aware that these images can serve as a signature, so each image has a unique hash.

Future work will focus on the forensic analysis of tools used to generate malicious MS documents and deobfuscation methods. To this end, tools like \texttt{Evil Clippy}\footnote{\url{https://github.com/outflanknl/EvilClippy}} and \texttt{LuckyStrike}\footnote{\url{https://github.com/curi0usJack/luckystrike}} will be examined, to determine their use in malicious campaigns. Moreover, we plan to examine further the attack surface that can be provided by an MS Office document as, e.g. the use of remote resources from XML has already been used to leak information, without even opening the file\footnote{\url{https://medium.com/@curtbraz/getting-malicious-office-documents-to-fire-with-protected-view-4de18668c386}}.

\section*{Acknowledgements}
This work was supported by the European Commission under the Horizon 2020 Programme (H2020), as part of the projects \textit{CyberSec4Europe} (Grant Agreement no. 830929) and \textit{LOCARD} (Grant Agreement no. 832735).

The content of this article does not reflect the official opinion of the European Union. Responsibility for the information and views expressed therein lies entirely with the authors.

\bibliographystyle{plain}
\bibliography{references}
\end{document}